# 274-GHz CMOS Signal Generator with an On-Chip Patch Antenna in a QFN Package

P. R. Byreddy, Z. Chen, H. S. Bakshi, W. Choi, A. Blanchard and K. K. O

A 274-GHz signal generator with an on-chip antenna in a quad-flat no-leads (QFN) package is demonstrated. The circuit is fabricated in a 65-nm Complementary Metal Oxide Semiconductor (CMOS) process. The effective isotropic radiated power (EIRP) measured from the packaged signal generator is −12.8-dBm at 274GHz with DC power consumption of 26.3mW. The packaging material can help improve the antenna performance, and integrated circuits with on-chip antennas operating at ~300GHz can be packaged using low-cost techniques.

*Introduction:* Capable Complementary Metal Oxide Semiconductor (CMOS) imaging, spectroscopy and communication circuits and systems operating at 200GHz and above have been reported [1-4]. Many of these systems use on-chip antennas for signal reception and transmission, and the effects of packaging [5], especially of affordable types of packaging have been of major concern. Despite this, no CMOS circuits in a conventional low-cost package operating in this frequency range have been reported. The work closest to addressing this concern is a 60-GHz FMCW radar encapsulated in a plastic package [6]. To bound the extent of this packaging concern and challenges, a 274-GHz CMOS signal generator with an on-chip patch antenna is packaged in a conventional low-cost quad-flat no-lead (QFN) package and its performance has been characterized. The packaged circuit delivers an effective isotropic radiated power (EIRP) of −12.8dBm at 274GHz while consuming 26.3mW. The measured EIRP is only 0.8dB lower than the simulated results without the packaging at the target operating frequency of 300GHz, despite ~10-% downshift in the operation frequency. 3-D electromagnetic (EM) simulations suggest that having a packaging layer over the patch can indeed improve the performance. These indicate that it should be possible to package integrated circuits with an on-chip antenna operating around 300GHz using low cost packaging techniques.

*Signal generator design:* Fig. 1*a* shows the schematic and Fig. 1*b* shows the die photograph of the signal generation circuit. The circuit is fabricated in a 65-nm CMOS process which has one polysilicon, ten copper, and one aluminium layers including the ~3-µm-thick top copper layer. The circuit is similar to that in [7] except it includes a nested inductor structure to reduce its area and to improve the phase noise. The circuit consists of an on-chip patch antenna and a push-push voltage-controlled oscillator (VCO). The measured output frequency is down shifted by 26GHz (less than 10%) from the 300-GHz target. The 2nd harmonic 274-GHz signal is extracted from the common mode node of the VCO which oscillates at a fundamental frequency of 137GHz [8]. Two cross-coupled pairs are stacked in series to form the VCO. The stacked cross-coupled pair topology improves negative-$g_m$ of the tank and allows a larger voltage swing [7,9]. The top cross-coupled pair, along with an inductor, $L_D$, oscillates around 135.5-137.8GHz whereas the bottom cross-coupled pair in combination with $L_S$ forms an oscillator operating around 45-47GHz [7]. The top oscillator (137GHz) is 3rd-order sub-harmonic injection-locked by the bottom oscillator (45.67GHz). The bottom oscillator is stabilized by injection-locking it with an external 22.83-GHz signal using an on-chip frequency doubler. A critical difference from [7] is that the inductor of the bottom oscillator, $L_S$, is folded up around the inductor of the top oscillator, $L_D$, so that they form a nested structure. Due to this nesting, both the inductors are coupled to each other with a coupling coefficient, $k$ of 0.17, which reduces the core area of the circuit in Fig. 1*b* by 35% and improves the phase noise by 2dB in simulations. The entire chain from the injection input at 22.83GHz to the push-push output of VCO at 274GHz has a frequency multiplication factor of 12. The antenna is formed using a half-wavelength patch using the ~1.2-µm-thick aluminium bond-pad layer, which is separated by about ~8µm from the ground plane formed by shunting the first six metal layers. The simulated radiation efficiency and gain of antenna are 34.6% and 0.8dB at 274GHz. The chip area including bond pads for DC and 22.83-GHz input is 450×900 µm².

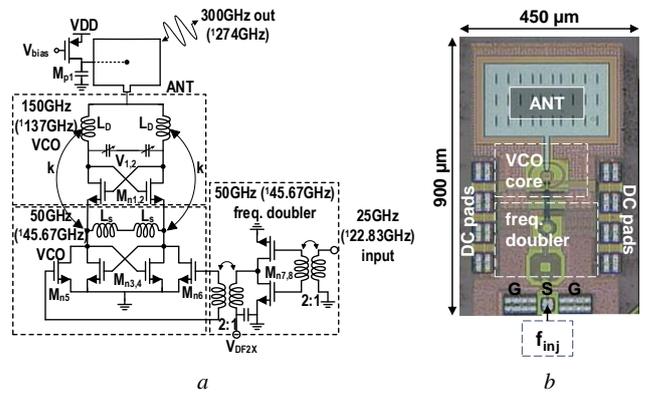

**Fig. 1** *Schematic and die photograph of signal generator*
*a* Circuit schematic ([1]Actual operating frequency)
*b* Die photograph

*Packaging:* The signal generator is packaged in a low-cost QFN package. The size of the package is 4×4mm² with 16 pads. The thickness of packaging material over the die is ~400µm. The material properties at the operating frequency are not available. The package is mounted on a 2-layer printed circuit board (PCB) using a low-cost FR-5 substrate as shown in Fig. 2*a*. The silicon chip is wire-bonded using gold wires to the pads of the QFN package.

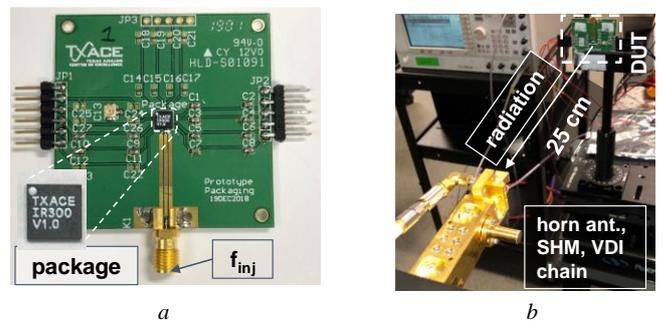

**Fig. 2** *Device under test (DUT) and measurement setup.*
*a* PCB with a signal generator in a low-cost QFN package
*b* Measurement setup

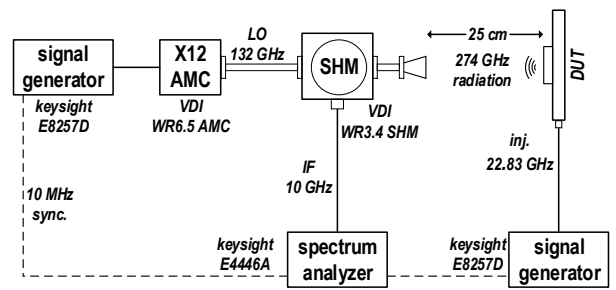

**Fig. 3** *Measurement setup for the packaged signal generator.*

*Measurement results:* Fig. 2*b* shows the photograph and Fig. 3 shows the schematic of the measurement setup for the packaged signal generator. The EIRP is measured using a VDI WR3.4SHM sub-harmonic mixer (SHM) connected to a 24-dB rectangular horn antenna. The LO to the SHM is provided using a VDI WR6.5AMC (Amplifier / Multiplier Chain) which has a multiplication factor of 12. The gain of the SHM is characterized using an OML S03MS-AG (X18 AMC) source module, whose power level is calibrated using a VDI PM4 power meter. An external signal at $f_{LO}/12$ is applied through a grounded co-planar waveguide (GCPW) on the PCB to injection lock the signal generator. The separation between the device under test (DUT) and horn antenna is set at 25cm to ensure $1/r^2$ dependence of the detected power. The measured output frequency was 274GHz, which is ~10% lower than the design target frequency of 300GHz due to uncertainties in the estimation of the parasitics associated with coupling of the inductors ($L_D$ and $L_S$) and the effects of packaging material on the oscillator core. To accommodate this shift, the pixel is injection-locked to a signal at ~22.83GHz. At an injection power level of 2dBm, the signal generation circuit exhibits a locking range of 700MHz at the output with a 2.3-GHz frequency tuning range. The EIRP of the signal generator estimated from the measurements using the Friis transmission equation is −12.8dBm. Fig. 4 shows the output spectrum of both free running and injection-locked signals. The total chip consumes 26.3mW of power while the oscillator and 3$^{rd}$ order sub-harmonic injection locking circuit together consume 21.6mW at $V_{DD}$=1.6V, and the frequency doubler dissipates 4.7mW at $V_{DD}$=1.0V.

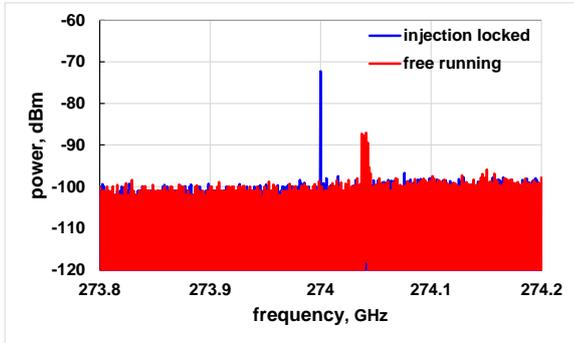

**Fig. 4** *Measured output spectrum of the packaged structure.*

Table 1 summarizes the measured and simulated performance and characteristics of the signal generation circuit. The measured EIRP is close to that expected from the simulations at 300GHz. To check whether this is possible, the patch antenna with an overcoat of 400-μm thick material with a loss tangent of 0.04 and a relative permittivity, $\varepsilon_r$ of 3.5 is simulated and summarized in Table 2. The loss tangent of 0.04 is significantly higher than the terahertz packaging materials listed in [5]. Despite this, the addition of overcoat improves the efficiency, match and gain at 274GHz and the 300-GHz target operating frequency, respectively. This may be caused by the overcoat working as a form of superstrate which improves the radiation characteristics by enhancing antenna to free space matching as well as the impedance matching between the antenna and circuit.

**Table 1:** Simulated (without packaging material) and measured (with packaging) performance of the injection locked signal generator.
[1]Effective isotropic radiated power [2]$\eta_{rad}$= radiation efficiency

|  | Simulation | **Measured** |
|---|---|---|
| Freq. (GHz) | 300 | **274** |
| Process | 65-nm CMOS | **65-nm CMOS** |
| Harmonic | 2 | **2** |
| $f_{in}$ (GHz) | 25 | **22.83** |
| [1]EIRP (dBm) | -12 | **-12.8** |
| Area (mm$^2$) | 0.45x0.9 | **0.45x0.9** |
| $P_{DC}$ (mW) | 24 | **26.3** |
| $|S_{11}|$ (dB) | -5.7 | - |
| [2]$\eta_{rad}$ (%) | 45.5 | - |
| Gain (dB) | 1.2 | - |
| Packaging | None | **QFN** |

**Table 2:** Simulation results of patch antenna with a 400-μm thick packaging material on top. Loss tangent and $\varepsilon_r$ are 0.04 and 3.5, respectively.
$\eta_{rad}$= radiation efficiency

| Parameter | No package | Package |
|---|---|---|
| $|S_{11}|$ at 300GHz (dB) | -5.7 | -8.2 |
| $|S_{11}|$ at 274GHz (dB) | -0.7 | -2.2 |
| $\eta_{rad}$ at 300GHz (%) | 45.5 | 46.7 |
| $\eta_{rad}$ at 274GHz (%) | 34.6 | 45.1 |
| Gain at 300GHz (dB) | 1.2 | 1.5 |
| Gain at 274GHz (dB) | 0.8 | 1.1 |

*Conclusion:* A 274-GHz CMOS signal generator with an on-chip patch antenna packaged in a low-cost QFN package is reported. The measured EIRP at 274GHz is only 0.8dB worse than the simulated results without the packaging at the target operating frequency of 300GHz. 3-D EM simulations suggest that having a packaging layer over the patch can actually improve the radiation efficiency even with a loss tangent of 0.04. The observed improvement in measurements over simulation may be due to the improvement of antenna efficiency and matching in the package. These results indicate that it should be possible to package integrated circuits with an on-chip antenna operating around 300GHz using low cost packaging techniques.

*Acknowledgments:* This material is based upon work supported by TI FTRC on Millimeter Wave and High Frequency Microsystems, the SRCco., JUMP and DARPA, and The Regents of the University of California.

P. R. Byreddy, Z. Chen, H. S. Bakshi, A. Blanchard, K. K. O (*Texas Analog Centre of Excellence and Department of ECE, The University of Texas at Dallas, 800 W. Campbell Rd, Richardson, TX 75080, USA*)

E-mail: pxb142130@utdallas.edu

W. Choi (*School of Electrical and Computer Engineering, Oklahoma State University, Stillwater, OK 74078, USA*)